\documentclass[english]{article}
\usepackage[LGR,T1]{fontenc}
\usepackage[latin9]{inputenc}
\usepackage{geometry}
\usepackage{color}
\usepackage{array}
\usepackage{amstext}

\makeatletter

\DeclareRobustCommand{\greektext}{%
  \fontencoding{LGR}\selectfont\def\encodingdefault{LGR}}
\DeclareRobustCommand{\textgreek}[1]{\leavevmode{\greektext #1}}
\DeclareFontEncoding{LGR}{}{}
\DeclareTextSymbol{\~}{LGR}{126}
\newcommand{\lyxmathsym}[1]{\ifmmode\begingroup\def\b@ld{bold}
  \text{\ifx\math@version\b@ld\bfseries\fi#1}\endgroup\else#1\fi}

\providecommand{\tabularnewline}{\\}

\usepackage[square, numbers, comma, sort&compress]{natbib}

\makeatother

\usepackage{babel}
\begin{document}

\title{Analysis and improvement of Tian-Zhang-Li voting protocol based on
controlled quantum teleportation}

\author{Kishore Thapliyal$^{a}$, Rishi Dutt Sharma$^{b}$ and Anirban Pathak$^{a}$}

\maketitle
\begin{center}
$^{a}$Jaypee Institute of Information Technology, A-10, Sector-62,
Noida, UP-201307
\par\end{center}

\begin{center}
$^{b}$National Institute Technology Patna, Ashok Rajhpath, Patna,
Bihar 800005, India
\par\end{center}
\begin{abstract}
Recently Tian, Zhang and Li (TZL) have proposed a protocol for voting
based on controlled quantum teleportation (Int. J. Theor. Phys. DOI
10.1007/s10773-015-2868-8). We have critically analyzed the protocol
and have shown that it's neither efficient nor secure. Further, it
is shown that in the TZL protocol, the scrutineer Charlie does not
have the required control over the voting process. Apart from showing
the limitations of TZL protocol, two improved protocols for quantum
voting along the line of TZL protocol are proposed here. One of the
proposed protocols is designed using a standard scheme of controlled
deterministic secure quantum communication, and the other one is designed
using the idea of quantum cryptographic switch which uses a technique
known as permutation of particles (PoP). A few possible alternative
approaches to accomplish the same task have also been discussed. Further,
the efficiencies of the proposed protocols are reported, and it is
shown that the proposed protocols are free from the limitations of
the TZL protocol, and they are more efficient than the TZL protocol.
\end{abstract}
Keywords: quantum voting, secure quantum communication, quantum cryptography,
TZL protocol, controlled secure quantum communication

\section{Introduction}

Since the pioneering work of Bennett and Brassard on quantum key distribution
(QKD) protocol \cite{bb84} (BB84 protocol), several protocols for
secure quantum communication have been proposed (for a detail review
see Chapter 8 of Ref. \cite{my-book}). Initial studies on secure
quantum communication were restricted to single particle based QKD
\cite{bb84,b92,vaidman-goldenberg}, entangled-state-based QKD \cite{ekert,Long-and-Liu},
quantum secure direct communication (QSDC) \cite{ping-pong,lm05},
deterministic secure quantum communication (DSQC) \cite{dsqc_summation,dsqcqithout-max-entanglement,dsqcwithteleporta,entanglement-swapping,reordering1,the:cao-and-song,the:high-capacity-wstate,With-Anindita-pla},
etc. However, in the last couple of years several innovative applications
of secure quantum communication have been proposed \cite{auction,e-commerce,qd,Hillery06pla-voting}.
These recently proposed schemes of secure quantum communications (some
of which are also hybrid in nature) have very interesting and important
applications in our daily life. Specifically, we would like to note
that in \cite{qd}, it is shown that protocols of quantum dialogue
can be modified to provide solution of the socialist millionaire problem;
innovative quantum solutions have also been provided for auction \cite{auction},
e-commerce \cite{e-commerce} and voting \cite{Hillery06pla-voting,TZL-voting,Vacaro-pra07-voting,li-08-voting,Horoshiko-PLA-11-voting,Hillery-pra-11-voting,jiang12-pra-voting,voting12}. 

The existing protocols for quantum communication can be broadly categorized
as: (A) insecure protocols (or protocols where security is not required,
like teleportation), and (B) secure protocols (like QKD). Insecure
protocols do not require any security measures while they may be used
in designing secure protocols for which security analysis is considered.
Teleportation, remote state preparation (RSP) and other variants of
teleportation fall under the Category A, whereas protocols for QKD,
QSDC and DSQC, etc., lie in Category B. The present paper is focused
on quantum voting, which is a particular type of protocol for secure
quantum communication. The reason for this specific choice is justified
as among the above mentioned interesting applications of secure quantum
communication, voting is of specific importance to our daily life.
Specifically, major policies of a state (nation) are decided on the
basis of voting, and in a democratic country policy makers are elected
by voting. In voting, privacy of the voter is of major importance
and as schemes of secure quantum communication can be made unconditionally
secure, it is natural to exploit the findings of the secure quantum
communication or cryptography for the purpose of designing schemes
for quantum voting. In 2006, Hillery et al., proposed the first scheme
of quantum voting \cite{Hillery06pla-voting}. Almost immediately
after that Vaccaro, Spring, and Chefles proposed an alternative scheme
for quantum voting \cite{Vacaro-pra07-voting}. Later on in 2011,
Hillery et al., proposed another scheme of quantum voting \cite{Hillery-pra-11-voting}
in which they circumvented some of the limitations of their earlier
scheme. Subsequently, several other protocols for quantum voting appeared
(\cite{jiang12-pra-voting,voting12} and references therein). All
these schemes were aimed to satisfy (i) security, (ii) verifiability,
and (iii) privacy. The importance of voting in public life motivated
us to analyze one of the most recent proposals for quantum voting.
To be precise, in what follows, we critically analyze a quantum voting
scheme recently proposed by Tian, Zhang and Li \cite{TZL-voting}
using a scheme of controlled quantum teleportation. Our analysis reveals
several limitations of the TZL protocol. In fact, it indicates that
controlled quantum teleportation is not the correct choice for implementing
a scheme of quantum voting, rather an analogous scheme based on controlled
secure quantum communication can perform the task more appropriately
and efficiently. To emphasize on this point, we have proposed two
new protocols for quantum voting which are based on two schemes of
controlled DSQC (CDSQC), and are free from the limitations of TZL
protocol. The first protocol uses a standard scheme of CDSQC. However,
the second protocol intrinsically uses a CDSQC scheme introduced recently
\cite{AP-cdsqc} as a variant of quantum cryptographic switch \cite{switch}.
It may be noted that the concept of quantum cryptographic switch was
introduced by Srinatha et al., in 2014 \cite{switch}, to describe
a CDSQC scheme where the controller can control the information the
receiver can extract to a continuously varying degree. In the cryptographic
switch and its variants (various schemes of controlled communication
discussed in Refs. \cite{AP-cdsqc,crypt-switch}) a technique known
as permutation of particles (PoP) plays a crucial role \cite{PoP}.
Specially, use of PoP reduces the requirement of quantum resources
and provides security. This interesting technique which was introduced
by Deng and Long in 2003 for QKD \cite{PoP} is used here to design
our second protocol which is more efficient than the TZL protocol
and the other protocol proposed here using the standard CDSQC scheme.
Further, it is shown that the proposed schemes are a bit general in
nature, as CDSQC can be implemented in various ways and each of those
different approaches can provide a scheme for quantum voting.

The remaining part of the paper is organized as follows. In Sec. \ref{sec:TZL scheme},
we briefly describe TZL scheme of quantum voting. In Sec. \ref{sec:Limitations-of-the TZL protocol},
limitations of TZL scheme are discussed. In Sec. \ref{sec:Improved-protocols},
we propose two improved schemes for quantum voting which are free
from the limitations of TZL protocol, and we also discuss a few alternative
approaches that can be used to obtain schemes of quantum voting analogous
to the TZL scheme. Finally, the paper is concluded in Sec. \ref{sec:Conclusion}.

\section{Controlled quantum teleportation and voting schemes of Tian, Zhang
and Li \label{sec:TZL scheme}}

In this section, we briefly describe the controlled quantum operation
teleportation scheme and the quantum voting protocol proposed in the
recent paper of Tian, Zhang and Li \cite{TZL-voting}.

\subsection{TZL scheme of controlled quantum operation teleportation\label{sub:TZL-ct}}

The TZL controlled teleportation scheme uses four qubit entangled
state which can be written as $\frac{1}{\sqrt{2}}\left(|GHZ^{0+}\rangle|0\rangle+\right.$
$\left.|GHZ^{1+}\rangle|1\rangle\right)$, where $|GHZ^{a\pm}\rangle=\frac{1}{\sqrt{2}}\left(|ijk\rangle\pm|\bar{i}\bar{j}\bar{k}\rangle\right)$
is a $GHZ$ state with $a$ as the decimal value of the binary number
$ijk$ for $i,j,k\in\left\{ 0,1\right\} $. The channel is shared
between three parties Alice, Bob and Charlie as follows: Alice possesses
the first two particles, whereas the third and last particles are
with Bob and Charlie, respectively. Further, Alice owns a quantum
state to be teleported. The scheme used in TZL quantum voting protocol
can be summarized in the following steps:
\begin{description}
\item [{Step~1:}] Alice performs a measurement in $GHZ$-basis on all
the three qubits of her possession.
\item [{Step~2:}] Alice announces her measurement outcome.
\item [{Step~3:}] Charlie performs a measurement in the computational
basis on his qubit and announces the outcome only when he wishes Alice
and Bob to perform the teleportation.
\item [{Step~4:}] Using Alice's and Charlie's measurement outcomes, Bob
performs a particular Pauli operation on his qubit using Table \ref{tab:TZL-table}
(which is same as Table 1 of Ref. \cite{TZL-voting}, except the fact
that Columns 3 and 4 are added here to illustrate a limitation of
TZL scheme)\textcolor{blue}{{} }to reconstruct the quantum state sent
by Alice.
\end{description}
It may be observed from Table \ref{tab:TZL-table} that if Alice announces
a measurement outcome $|GHZ^{0+}\rangle$ then for Charlie's measurement
outcome of $|0\rangle$ ($|1\rangle$) Bob needs to apply $I$ ($X$)
operation. This was what proposed in TZL scheme. However, a closer
look into the table (see Columns 3 and 4 of Table \ref{tab:TZL-table})
reveals that\textcolor{blue}{{} }Bob's measurement outcome is independent
of Charlie's measurement outcome. Specifically, if we neglect the
global phase, then even before the application of Bob's unitary operation,
the quantum state of Bob's qubit is the same for both the outcomes
of Charlie. Further, it is worth noting that an application of $X$
gate on $|+\rangle$ or $|-\rangle$ leaves the state unchanged up
to the global phase. Hence, it can be concluded that Charlie's measurement
outcome is irrelevant for Bob and even before Charlie performs measurement
on his qubit, Bob can reconstruct Alice's teleported state only by
knowing the outcome of Alice's measurement. This fact can be mathematically
verified by writing density matrix for the combined states of Charlie
and Bob (after Alice's measurement) and subsequently tracing out Charlie's
qubit, and measuring Bob's qubit in $\left\{ |+\rangle,|-\rangle\right\} $
basis. Similarly, for Alice's measurement outcome $|GHZ^{3+}\rangle,\,|GHZ^{3-}\rangle$
and $|GHZ^{0-}\rangle$ Bob can apply $I,\, Z$ and $Z$ operations,
respectively, to obtain the state teleported by Alice, without bothering
to know the measurement outcome of Charlie. Therefore, Charlie's control
is missing in the TZL controlled teleportation scheme if the sender
wishes to teleport only $|+\rangle$ or $|-\rangle$ state. 

\begin{table}
\begin{centering}
\begin{tabular}{|>{\centering}p{3cm}|>{\centering}p{3cm}|>{\centering}p{2cm}|>{\centering}p{2cm}|>{\centering}p{2cm}|}
\hline 
Alice's measurement result & Charlie's measurement result & Bob's qubit before applying unitary operation

(when initial state is $|+\rangle$) & Bob's qubit before applying unitary operation

(when initial state is $|-\rangle$) & Bob's unitary operation\tabularnewline
\hline 
$|GHZ^{0+}\rangle$ & $|0\rangle$ & $|+\rangle$ & $|-\rangle$ & $I$\tabularnewline
\hline 
 & $|1\rangle$ & $|+\rangle$ & $-|-\rangle$ & $X$\tabularnewline
\hline 
$|GHZ^{0-}\rangle$ & $|0\rangle$ & $|-\rangle$ & $|+\rangle$ & $Z$\tabularnewline
\hline 
 & $|1\rangle$ & $-|-\rangle$ & $|+\rangle$ & $iY$\tabularnewline
\hline 
$|GHZ^{3+}\rangle$ & $|0\rangle$ & $|+\rangle$ & $-|-\rangle$ & $X$\tabularnewline
\hline 
 & $|1\rangle$ & $|+\rangle$ & $|-\rangle$ & $I$\tabularnewline
\hline 
$|GHZ^{3-}\rangle$ & $|0\rangle$ & $-|-\rangle$ & $|+\rangle$ & $iY$\tabularnewline
\hline 
 & $|1\rangle$ & $|-\rangle$ & $|+\rangle$ & $Z$\tabularnewline
\hline 
\end{tabular}
\par\end{centering}

\caption{\label{tab:TZL-table} The relationship between the measurement outcomes
of Alice and Charlie with Bob's operation in TZL scheme. Here, $GHZ^{0\pm}=\frac{\left(|000\rangle\pm|111\rangle\right)}{\sqrt{2}}$
and $GHZ^{3\pm}=\frac{\left(|011\rangle\pm|100\rangle\right)}{\sqrt{2}}$. }
\end{table}

\subsection{TZL scheme of quantum voting}

The TZL voting scheme has three phases: Initial phase, voting phase
and counting phase. The scheme begins with the initial phase in which
the notary organization CA (a trusted third party) issues quantum
IDs to all eligible voters, scrutineer (controller) and tallyman.
Finally, it ends with the counting phase where the controller decrypts
and announces the voting rules of each voter, using which the tallyman
decides the number of votes in favor. In between the most important
phase of the scheme (i.e., voting scheme) happens in the following
steps.
\begin{description}
\item [{Voting~1:}] The controller Charlie sets up a bulletin board, i.e.,
a classical channel to make the announcements.
\item [{Voting~2:}] The tallyman Bob authenticates identity of the voters
with the help of the trusted third party CA. After that, Bob sends
a qubit in $|+\rangle=\frac{|0\rangle+|1\rangle}{\sqrt{2}}$ state
to all the authenticated voters. 
\item [{Voting~3:}] Charlie also authenticates the identity of the voters
taking CA's help. Further, Charlie sends the voting rule to $i$th
voter ${\rm Alice_{i}}$ using BB84 encryption after knowing her serial
number. For example, Charlie can assign $i$th voter quantum operations
$I$ and $Z$ gates to encode ``yes'' and ``no'', respectively.
The only other possibility of encoding scheme for Charlie will be
to use $I$ ($Z$) gate for sending ``no'' (``yes''). \\
Here, it is important to note that the voting rule is known to $i$th
voter and Charlie only%
\footnote{The purpose of keeping the voting rule secret is two fold. Firstly,
it provides controller/scrutineer Charlie a control over the time
at which the votes will be counted. In other words, until Charlie
discloses the voting rules, the tallyman Bob will not be able to count
votes. Secondly, when voting happens sequentially, it restricts $i+1$th
voter from knowing the trend of voting happened until then (i.e.,
the trend up to the time $i$th voter has voted). %
}.
\item [{Voting~4:}] At the voter's request Bob prepares a four qubit entangled
state and sends first two qubits to the voter, while the last one
to Charlie, keeping the third qubit with himself.
\item [{Voting~5:}] To cast her vote the $i$th voter ${\rm Alice}_{i}$
will encode her vote (voting information) on the qubit sent by Bob.
The encoding is to be done according to the voting rule assigned to
her by Charlie in Voting 2. Then Alice performs a measurement in GHZ-state
basis on the encoded qubit and two qubits of the quantum channel set
by Bob in Voting 4. This is followed by an announcement of the measurement
outcome on the bulletin board prepared in Voting 1 and Charlie's measurement
of his qubit of the shared channel (in Voting 4) in computational
basis. Finally, Charlie also announces the measurement outcome.
\item [{Voting~6:}] Using the measurement outcomes of Alice and Charlie,
Bob can perform suitable Pauli operator (from Table \ref{tab:TZL-table})
to obtain the teleported quantum state. Finally, the reconstructed
state is measured in diagonal basis and the result is broadcasted
classically. 
\item [{Voting~7:}] In a similar approach, Bob performs the voting procedure
for all the voters.
\end{description}
In Voting 5 and Voting 6 the scheme of controlled teleportation described
in Sec. \ref{sub:TZL-ct} has been used and as a consequence, this
scheme of quantum voting may be viewed as a quantum voting scheme
based on controlled quantum teleportation and in the TZL paper it
was referred in this manner.

\section{Limitations of the TZL protocol \label{sec:Limitations-of-the TZL protocol}}

A careful look into the TZL protocol for quantum operation teleportation
reveals that Alice actually encodes a single bit of information and
transmits that to Bob. Say, we consider encoding by operation of identity
is ``0'' and encoding by operation of $Z$ as ``1''. Secure controlled
quantum communication of 1 bit of classical information can be performed
in a large number of ways and controlled teleportation is neither
essential nor a good choice. This is not a good choice because, teleportation
or controlled teleportation in its original form is not secure, and
standard controlled teleportation schemes require $n$-partite entanglement
($n\geq3)$ and at least 2 bits of classical communication to communicate
one bit of voting information. There are many alternative ways to
perform the same task and controlled teleportation is just one of
the ways and it's not secure in its original form. Further, in TZL
protocol, Alice always encodes her message on the quantum state $|\phi_{a}\rangle=\frac{1}{\sqrt{2}}\left(|0\rangle+|1\rangle\right)=|+\rangle$,
so it is not required that Bob prepares the state. Alice herself can
prepare the state. In fact, when Bob prepares this quantum state and
transmits that to Alice via a quantum channel that just reduces the
quantum efficiency of the protocol as it increases the amount of quantum
communication. Further, in practical situations (i.e., in a noisy
environment) this additional communication would reduce the fidelity
of the final state. We can further note that after encoding operation
of Alice, she knows which state is to be teleported (it is either
$|+\rangle$ or $|-\rangle$ depending upon the operator used by her)
as she knows what operation she has performed on the initial state
$|+\rangle$. Teleportation of a known state is technically referred
to as the remote state preparation, and thus the problem here is of
controlled remote state preparation. It is well known that quantum
channel that can perform (controlled) teleportation can always perform
(controlled) remote state preparation. However, for RSP we don't need
to prepare an additional qubit in $|+\rangle$ state. Further, controlled
quantum teleportation and controlled RSP of one qubit can be performed
using 3 qubit states only. 4 qubit state is not required. For example,
in Ref. \cite{ghz-like-comment,Anindita}, we have already shown how
controlled teleportation can be performed using GHZ-like states. We
note this point just to show that use of 4-qubit state in TZL protocol
was unnecessary. However, we are not going to elaborate a 3-qubit
controlled teleportation based version of TZL quantum voting scheme
as the other limitations of the TZL protocol will remain present in
that, too. 

Another major limitation of the TZL voting scheme lies in the fact
that in Voting 4 step, Bob prepares the quantum state to be used for
controlled teleportation. Here, Bob can always prepare a quantum state
such that Charlie's state is separable from the state shared by Alice
and Bob, and this would help him to circumvent Charlie's control without
being detected. To be precise, consider that Bob randomly prepares
one of the states: $|GHZ^{0+}{\rm \rangle_{A_{1}A_{2}B}|0\rangle_{C}=}\frac{1}{\sqrt{2}}\left(|000\rangle+|111\rangle\right)_{{\rm A_{1}A_{2}B}}|0\rangle_{{\rm C}}$
or, $|GHZ^{0+}{\rm \rangle_{A_{1}A_{2}B}|1\rangle_{C}=}\frac{1}{\sqrt{2}}\left(|000\rangle+|111\rangle\right)_{{\rm A_{1}A_{2}B}}|1\rangle_{{\rm C}}$
and shares the state as described in the step Voting 4. In this case,
Charlie's measurement outcome will be randomly $|0\rangle$ and $|1\rangle$,
but it will not have any control over the voting scheme. Thus, the
fact that in TZL protocol controller Charlie does not have the desired
control over the voting process is now established in two different
ways. 

It is easy to circumvent the problem arising due to the fact that
Bob has been assigned the responsibility to prepare the quantum channel.
Specifically, this problem can be circumvented by allotting the state
preparation task to Charlie. However, even this modification of TZL
protocol would not solve the problem completely (i.e. would not provide
the required control to scrutineer). Specifically, as mentioned in
Sec. \ref{sec:TZL scheme}, Charlie's control in the protocol will
be missing as Bob can reconstruct the state even before Charlie's
measurement and announcement of corresponding outcome. 

Further, in the TZL protocol, it is intrinsically assumed that the
sender sends a qubit to the receiver and there is no eavesdropper
who can attempt to replace the original qubit with an auxiliary one
and finally change the vote with her choice. This is not desired.
In fact, the voters would come to know about this attack when their
votes are announced on the bulletin board. It is reasonable to think
that it would be hard to claim unfairness as the voter would have
to disclose her original choice in that case. This can be circumvented
by using decoy qubits while sharing the qubits and quantum channel.
Using a proper eavesdropping checking scheme this kind of an attack
can be circumvented.

\section{Improved protocols\label{sec:Improved-protocols}}

We have already shown that in the TZL protocol, the task performed
is essentially communication of one bit of information by controlled
RSP of a qubit. This observation may motivate one to modify the original
TZL protocol by replacing the controlled teleportation scheme used
in TZL protocol by a scheme of controlled RSP. However, such a modification
will not succeed to circumvent the limitations of the original TZL
scheme. In fact, in such a modified scheme, Alice's vote will be leaked
as soon as she will disclose the outcome of her measurement. Technically,
here the ignorance of tallyman Bob or an eavesdropper is of one bit
(they only need to know whether the state to be prepared remotely
is a $|+\rangle$ or a $|-\rangle$state), and Alice's disclosure
removes this ignorance (in other words, provides required one bit
information to both Bob and eavesdropper.  Thus, we note that to obtain
an improved protocol for quantum voting, it would be apt to use a
scheme of secure controlled communication (say a scheme of CDSQC)
over an insecure quantum communication protocol (say a scheme of controlled
teleportation or controlled RSP). Keeping these facts in mind, let
us now introduce our first scheme of quantum voting based on CDSQC.

\subsection{Protocol 1: Quantum voting based on controlled-deterministic secure
quantum communication (CDSQC) \label{sub:Protocol-1}}

The TZL quantum voting protocol can be visualized as a protocol for
transmitting one bit of classical information by teleporting a qubit
from the voter (sender) to a tallyman (receiver) under the supervision
of controller/scrutineer. As discussed above, the TZL protocol in
its original form has several limitations. However, the controlled
communication task performed by TZL protocol can be performed in a
secure manner by using various alternative approaches. To be precise,
there exist various protocols for controlled deterministic secure
quantum communication (CDSQC), which are controlled version of DSQC
or QSDC protocols  \cite{AP-cdsqc}. In what follows, we will show
that any CDSQC protocol can be reduced to a protocol of binary quantum
voting. To elaborate this point, let us first explicitly provide an
e-voting scheme based on a particular CDSQC protocol. The scheme proposed
below is designed along the line of TZL protocol, and the initial
phase and the counting phase of the TZL protocol remain unchanged
in our protocol. Only the voting phase is modified. Specifically,
the voting phase of the improved scheme can be summarized as follows:
\begin{description}
\item [{CDSQC~1:}] Same as Voting 1.
\item [{CDSQC~2:}] Bob only authenticates the identity of the voters with
the help of the trusted third party CA.
\item [{CDSQC~3:}] Same as Voting 3.
\item [{CDSQC~4:}] After receiving the voter's request, Charlie prepares
a three qubit entangled state ($GHZ$-like state) and sends first
and second qubits to the voter and tallyman, respectively. He keeps
the third qubit with himself.\\
 Here, it is important that the controller (Charlie) prepares the
quantum state and distribute it among the users, unlike Voting 4 where
the same task was done by Bob. Thus, the quantum channel prepared
by Charlie is 
\begin{equation}
|\psi\rangle_{GHZ-{\rm like}}=\frac{|\psi_{1}\rangle|0\rangle+|\psi_{2}\rangle|1\rangle}{\sqrt{2}},\label{eq:CDSQC-channel}
\end{equation}
where $|\psi_{i}\rangle\in\left\{ |\psi^{+}\rangle,|\psi^{-}\rangle,|\phi^{+}\rangle,|\phi^{-}\rangle\right\} $.
Here, $|\psi^{\pm}\rangle=\frac{|00\rangle\pm|11\rangle}{\sqrt{2}}$
and $|\phi^{\pm}\rangle=\frac{|01\rangle\pm|10\rangle}{\sqrt{2}}$
are Bell states. Further, a different choice of controller's qubit
in $\left\{ |+\rangle,|-\rangle\right\} $ basis would give us a $GHZ$
state. Therefore, GHZ states are also suitable for the proposed scheme%
\footnote{In fact, an arbitrary choice of controller's qubit in $\left\{ |a\rangle,|b\rangle\right\} $
basis will work, where the measurement outcome in this basis will
reveal the information about the quantum channel shared by Alice and
Bob. In brief, any quantum state of the form $|\psi\rangle_{GHZ-{\rm like}}=\frac{|\psi_{1}\rangle|a\rangle+|\psi_{2}\rangle|b\rangle}{\sqrt{2}}$
can be used to implement this protocol.%
}.
\item [{CDSQC~5:}] After receiving the first particle of the quantum channel
Alice operates $I$ or one of the Pauli gates ($X,\, iY$ or $Z$)
to cast her vote as ``yes'' or ``no'' as the voting rule set by
the controller in CDSQC 3. Then she sends the encoded qubit to Bob.
This is followed by Charlie's measurement of the third qubit in computational
basis and announcement of his measurement outcome. \\
It would be worth mentioning here that dense coding is not exploited
here. A scheme exploiting quantum dense coding can be designed where
the voters can encode 2 bits of information. 
\item [{CDSQC~6:}] Depending upon the measurement outcomes of Charlie,
Bob can perform Bell state measurement on the first two qubits and
broadcast the measurement outcome on the bulletin board.
\item [{CDSQC~7:}] In a similar approach, Bob performs the voting procedure
for all the voters.
\end{description}
For sharing the entangled channel in CDSQC 4 or sending the encoded
qubit in CDSQC 5 suitable decoy qubits are prepared and only after
going through a successful eavesdropping checking the next step is
followed. More detail of decoy qubit assisted eavesdropping checking
schemes and suitable choices of decoy qubits can be found in Ref.
\cite{decoy}.

To show that the proposed scheme is more efficient we use the quantitative
measure of efficiency as discussed in Refs.\textcolor{red}{{} }\cite{AP-cdsqc,defn-of-qubit-efficiency},
i.e., $\eta=\frac{c}{q+b},$ where $c$ denotes total classical bits
transmitted using the\textcolor{magenta}{{} }total number of qubits
$q$, while $b$ is the classical communication involved in decoding
of information (classical communication required for eavesdropping
check is not included)\textcolor{blue}{.} Usually, $\eta$ is referred
to as qubit efficiency and is frequently used to compare the protocols
of secure quantum communication. In the proposed protocol based on
CDSQC, for each vote (a bit of classical information) to be transmitted
a 3 qubit state is used. Additionally, 3 decoy qubits are required
to ensure security. Charlie informs Alice the voting rule with 1 bit
of classical communication which he announces in the counting phase
with another 1 bit of communication; and he also announces his measurement
outcome at the end with 1 bit of information. Thus, $q=6,\, b=3$
and $c=1$, and as a consequence qubit efficiency of the proposed
protocol is $\eta=11.11\%$. This is not high, but if we compare it
with the original TZL protocol we can recognize that qubit efficiency
of the protocol proposed here is greater than that of TZL protocol.
This point can be clearly understood if we note that for the original
TZL protocol, $q=5$, as a 4-qubit cluster state is used and as Bob
transmits another qubit on which voting is to be done. Further, $b=6,$
as the announcement of measurement outcomes of Alice and Charlie require
transmission of classical information of 3 bits and 1 bit, respectively,
in addition, disclosure of voting rules involves transmission of 2
bits (as it is done twice once from Charlie to Alice and once from
Charlie to Bob). Thus, for the original TZL protocol qubit efficiency
$\eta=\frac{1}{11}=9.09\%$ only. Here it would be apt to note that
the original TZL protocol is insecure, and since no qubit is used
as decoy qubit to check eavesdropping $q$ is relatively low. However,
if in the original TZL protocol Bob desires to check that the quantum
states prepared by him in Voting 2 and Voting 4 states reach the desired
users without eavesdropping he has to use 4 decoy qubits for each
vote, and that would make $q=9$. In that case, $\eta$ for TZL would
be $\frac{1}{15}=6.67\%,$ which only $60\%$ of the efficiency of
the protocol proposed here.

The qubit efficiency can be increased either by increasing the number
transmitted classical bits $c$ (however, we have to discard this
possibility as in case of binary voting described here $c$ is always
1) and/or decreasing $b$ and/or $q$. Increase in the amount of transmitted
classical bits $c$ using the same resource will allow us to construct
new schemes of voting (where more than two choices exist, say there
are four candidates and the voter has to choose one), which is beyond
the realm of binary voting discussed in the original TZL protocol
and the protocol proposed here. So in this paper, we don't discuss
the possibilities associated with the increase in $c$. We will discuss
this possibility elsewhere. In the protocol proposed here $q$ and
$b$ are already reduced with respect to TZL protocol. In the next
section, we will propose another scheme of quantum voting protocol
with further reduced values of $q$ and $b$. Before we describe the
new protocol, we would like to note that $b$ can be reduced by avoiding
different voting rules for each voter and using same voting rule for
all voters (as identity in favor and one of the 3 Pauli operators
for against). This would reduce the classical communication associated
with the protocol by 2 bits.

\subsection{Protocol 2: Quantum voting based on controlled-deterministic secure
quantum communication (CDSQC) solely using Bell states\label{sub:Protocol-2}}

In the previous section, we have proposed a scheme for quantum voting
using a GHZ-like state, which is three qubit maximally entangled state.
This reduced the requirement of quantum resources with respect to
the original TZL protocol, where 4-qubit entanglement was used. However,
tripartite entanglement is not essential. In this section, we will
provide a scheme for quantum voting using bipartite entanglement (Bell
states). This would increase efficiency of the voting protocol and
reduce the difficulties associated with the generation and management
of multi-partite entanglement. Keeping the initial phase unchanged
a modified scheme for quantum voting can be described as follows:
\begin{description}
\item [{CDSQC2~1:}] Same as CDSQC 1.
\item [{CDSQC2~2:}] Same as CDSQC 2.
\item [{CDSQC2~3:}] Charlie prepares $n$ copies of one of the Bell states.
He sends first qubit of the $ith$ Bell state to the $ith$ voter
and and prepares an $n$-qubit string with the second qubits of all
the Bell states. Then he applies a permutation operator on the string
formed using the second qubits and sends that to Bob. \\
In this step, security of transmitted qubits are ensured by random
insertion of an equal number of decoy qubits (in other words by testing
half of the travel qubits using BB84 subroutine or GV subroutine).
Iff the errors obtained by a standard eavesdropping checking process
is found below the tolerable limit they proceed to the next step%
\footnote{Thus, possibilities of eavesdropping are checked using standard decoy
qubit based techniques, like BB84 subroutine or GV subroutine (for
detail of these eavesdropping checking techniques see our earlier
works \cite{my-book,AP-cdsqc,crypt-switch,decoy}). A suitable choice
of decoy qubits and the subroutine can be opted depending on the interaction
with the environment or the nature of noise present in the channel
(for a detail discussion see Ref. \cite{decoy}).%
}.
\item [{CDSQC2~4:}] Each voter can encode her vote on the qubit received
by her using the known voting rule. The predecided voting rule can
be to operate $I$ for ``yes'' and one of the Pauli gates ($X,\, iY$
or $Z$) for ``no''. She sends the encoded qubit to Bob afterward.
\\
Exploiting dense coding capacity of the Bell states Alice could have
sent 2 bits of information. This can add some more features to the
voting process, which will be explored elsewhere. Again in this step
a decoy qubit based security subroutine should be employed.
\item [{CDSQC2~5:}] After all the voters have sent their qubits of Bell
states to Bob with their voting information he notifies Charlie.
\end{description}
This is followed by the counting phase in which Charlie announces
information regarding permutation operator and the initial state.
This allows Bob to find out which qubit of the sequence received by
him directly from Charlie was initially entangled to which qubit sent
by a voter. Thus, after the announcement of Charlie, Bob can perform
Bell measurement on partner particles to obtain the vote cast by each
voter as he already knows the initial Bell states. Finally, he counts
all the votes and announces the total number of votes in favor.

Similar to the protocol proposed in the last subsection quantitative
measure of efficiency \cite{AP-cdsqc} can be calculated. In this
scheme, for each vote (a bit of classical information) to be transmitted
2 qubits are used directly. In addition 3 decoy qubits are also required
to ensure security. Further, Charlie has to reveal an additional 1
bit of information for decoding the votes. Thus, $c=1,\, q=5,\, b=1$
and as a consequence, the efficiency of this PoP-based protocol of
quantum voting is $\eta=16.67\%$, which is higher than the previous
proposal and the original TZL protocol.

\subsection{Possible alternative approaches that can lead to schemes of quantum
voting}

Clearly, the quantum voting schemes described above are based on schemes
of CDSQC. As mentioned in Subsection \ref{sub:Protocol-1}, there
are various ways to implement schemes of CDSQC. Specifically, CDSQC
can be implemented using a large set of entangled states (e.g., W
state, $GHZ$ state, $GHZ$-like state, $Q_{4}$ state, $Q_{5}$ state,
cluster state, $|\lyxmathsym{\textgreek{W}}\rangle$ state, Brown
state, etc.) \cite{AP-cdsqc}. Though to exploit the dense coding
capacity of these states it would be preferable to use these states
for sending more information.

The quantum voting task can also be achieved using entanglement swapping
based CDSQC scheme. In this scheme a large number of copies of a 3-qubit
entangled state (\ref{eq:CDSQC-channel}) is prepared by the controller.
The qubits are shared between the voters and the tallyman, in such
a way that the first two qubits of a 3-qubit entangled state are sent
to one of the voters, while a sequence prepared by all the last qubits
is sent to the tallyman Bob after applying a permutation operator
on the sequence. Subsequently, the voters can follow entanglement
swapping based DSQC scheme proposed by Shukla et al., \cite{dsqc-ent-swap}.
Finally, the tallyman will require an additional information of the
permutation operator from the controller to retrieve the votes. It
is worth noting here that in this protocol the encoded qubits do not
travel through the channel \cite{AP-cdsqc,dsqc-ent-swap}.

It may be noted that except the voting part where security is ensured
by the security proofs of the CDSQC protocol used here to design the
voting protocols, the rest of the scheme is the same as the original
TZL protocol and hence the verifiability and privacy established in
TZL paper is applicable to our protocols, too. Keeping this in mind,
we have not elaborately discussed these two important aspects of voting
protocols in this paper.

\section{Conclusion\label{sec:Conclusion}}

The TZL quantum voting scheme has been critically analyzed here. Specifically,
it has been established that the TZL protocol is neither secure nor
efficient, and the scrutineer Charlie does not have the required control
over the voting process. Specifically, we noted that a quantum communication
scheme (controlled teleportation) where security measures are not
involved directly had been used to construct a protocol for performing
a highly classified job (voting). Clearly, it would have been more
appropriate to investigate the possibility of modifying a scheme of
secure quantum communication to build a scheme for quantum voting.
This approach is followed in the present paper to propose two new
schemes for secure and efficient quantum voting based on CDSQC. The
proposed schemes are shown to be free from the limitations of the
TZL protocol. Further, a few modified alternative approaches suitable
for the same task have also been discussed. In brief, limitations
of the TZL protocol have been explored and a set of alternatives has
been provided for the realization of binary quantum voting. We conclude
the paper with a hope that in the near future the experimentalists
will find it interesting to implement one or more of the CDSQC based
schemes for binary quantum voting discussed in this paper.

\end{document}